\documentclass[reprint,nofootinbib,amsmath,amssymb,fontenc aps]{revtex4-1}
\usepackage{graphicx,dcolumn,bm,hyperref,float}
\usepackage{anyfontsize}
\usepackage{color}
\usepackage{xcolor}

\newcommand{\beq}{\begin{equation}}
\newcommand{\eeq}{\end{equation}}
\newcommand{\bea}{\begin{eqnarray}}
\newcommand{\eea}{\end{eqnarray}}
\newcommand{\bef}{\begin{figure}}
\newcommand{\eef}{\end{figure}}

\newcommand{\tBH}{t_{\mbox{\tiny{BH}}}}
\newcommand{\tH}{t_{\mbox{\tiny{H}}}}
\newcommand{\Hub}{H_\Lambda}
\newcommand{\bb}{b}
\newcommand{\gam}{\gamma}
\newcommand{\mpl}{M_{\mbox{\tiny{Pl}}}}

\begin{document}

\title{\fontsize{11.9}{15}\selectfont Possible Relation between the Cosmological Constant and Standard Model Parameters}

\author{Mark P.~Hertzberg$^{1}$}
\email{mark.hertzberg@tufts.edu}
\author{Abraham Loeb$^2$}
\email{aloeb@cfa.harvard.edu}
\affiliation{$^1$Institute of Cosmology, Department of Physics and Astronomy, Tufts University, Medford, MA 02155, USA
\looseness=-1}
\affiliation{$^2$Department of Astronomy, Harvard University, 60 Garden Street, Cambridge, MA 02138, USA
\looseness=-1}
%\affiliation{$^3$Department of Physics, Harvard University, Cambridge, MA 02138, USA
%\looseness=-1}
%\affiliation{$^4$Department of Physics, Brown University, Providence, RI, 02912, USA
%\looseness=-1}

\begin{abstract}
We propose possible properties of quantum gravity in de Sitter space, and find that they relate the value of the cosmological constant to parameters of the Standard Model. In de Sitter space we suggest (i) that the most sharply defined observables are obtained by scattering objects from the horizon and back to the horizon and (ii) that black holes of discrete charge are well defined states in the theory. 
For a black hole of minimal discrete electric charge, we therefore demand that a scattering process involving the black hole and a probe can take place within a Hubble time before evaporating away, so that the state of a discretely charged black hole is well defined. By imposing that the black hole's charge is in principle detectable, which involves appreciably altering the state of a scattered electron, we derive a relation between the Hubble scale, or cosmological constant, and the electron's mass and charge and order one coefficients that describe our ignorance of the full microscopic theory. This gives the prediction $\Lambda\sim 10^{-123\pm2}\mpl^4$, which includes the observed value of dark energy. We suggest possible ways to test this proposal.
\end{abstract}

\maketitle

%\tableofcontents

\section{Introduction}
%{\em Introduction}.---
The values of the fundamental constants of the Standard Model and gravitation remain mysterious. There is currently no known principle that links any of their values.\footnote{Some previous interesting works to try to relate parameters, such as the weak scale to the Planck scale and forms of vacuum energy, includes Refs.~\cite{Arkani-Hamed:2000ifx,Chacko:2004ky,Wang:2018kly}. Our work will differ from these previous works in several ways; our stated principles, our usage of black holes, and the final relation between constants.}
The Standard Model consists of 26 parameters (if we include the massive neutrino sector and the strong sector's $\theta$-angle and work in natural units) and general relativity includes 1 additional parameter: the cosmological constant $\Lambda$ (plus Newton's constant, which we can use to set units). In terms of the Planck scale $\mpl$, the cosmological constant is observed to be fantastically small $\Lambda\approx 10^{-123}\,\mpl^4$ (in the convention of treating it as an energy density and assuming it is indeed the dark energy). There is currently no satisfactory explanation for its value; for some reviews, see Refs.~\cite{Weinberg:1988cp,Garriga:2000cv,Peebles:2002gy,Padmanabhan:2002ji,Frieman:2008sn}.
 To our knowledge, the only candidate dynamical explanation for its smallness (without appealing to anthropics, which may over-predict the value anyhow \cite{Loeb:2006en}) is an entropic argument; low $\Lambda$ implies large entropy of de Sitter space and hence there is an exponentially strong preference for small $\Lambda$ \cite{Hawking:1984hk}. This argument, however, overshoots and predicts $\Lambda$ should be effectively zero.

In this work, we suggest a possibility to improve the situation by appealing to hypothesized principles of quantum gravity. By imposing these principles on charged black holes, we derive a bound on the cosmological constant in terms of the charge and mass of the lightest particles. Interestingly, by saturating this bound (which can be motivated), we find that the predicted $\Lambda$ is roughly consistent with the observed value.

The outline of our paper is as follows: In Section \ref{Principles} we summarize our principles. In Section \ref{Relation} we show how this leads to a relation between fundamental constants. In Section \ref{Discussion} we discuss our findings and future work to test this proposal.

\section{Principles}\label{Principles}
%{\em Principles}.---

We suggest the following hypothetical principles that a theory of quantum gravity might impose upon the structure of black holes:
\begin{itemize}
\item There exists well defined states of black holes with any discrete charge (multiple of electron charge, $e$).
\item In de Sitter space, the sharpest defined observables arise from scattering from and to the horizon.
Hence, scattering processes must exist that can learn about the black hole's charge.
\end{itemize}
Here we are only referring to non-extremal black holes; there can be some large upper bound on their charge, but this will not be directly relevant.

The first principle is motivated by the observational fact that electric charge is discrete (quantized), and one may anticipate that this remains true in quantum gravity and that black holes can indeed carry this charge. The second principle is motivated by the fact that in standard relativistic quantum mechanics in asymptotically flat spacetime, the sharpest observables are scattering amplitudes from and to infinity. So for de Sitter space, which prevents such experiments at infinity, one may anticipate scattering from and to the horizon as the appropriate update.

Taken together, we can consider black holes of any discrete charge and enquire under what conditions a scattering experiment can be performed to clearly learn the black hole's charge. We take as a criterion that the scattering off a probe can give rise to an electric effect comparable to or larger than the gravitational effect. The most difficult amount of charge to detect is a black hole of minimal charge $e$, i.e., one electron or proton (or anti-particle).

In order to construct black hole states whose minimal charge can be measured, we can use the black hole's mass $M$ as a free parameter. If we consider black holes of extremely large mass $M$, their effects on scattering probes will be entirely dominated by the gravitational scattering. This is not useful to learn about the black hole's minimal charge. On the other hand, black holes of extremely small mass $M$ will Hawking evaporate away before the scattering process from and to the horizon is complete. So these too are not useful to learn about the black hole's minimal charge. (To be clear; we are not saying such light and therefore rapidly evaporating black holes cannot exist. We are only suggesting that quantum gravity will impose that there can exist heavier charged black hole states that are more sharply defined.)

Hence the optimal choice is a black hole that is as light as possible, but just heavy enough to survive the scattering process. The proposal is that these states should exist in principle, even if not easily produced. The duration of the full scattering process is controlled by the Hubble scale of de Sitter space, which is in turn controlled by the cosmological constant. As we will mention, depending on parameters, the black holes of interest may undergo rather rapid neutralization through the Schwinger effect, so the full process can involve absorption and emission of its charge. See Figure \ref{ScatteringDiagram} for a depiction of the full scattering process. 

Since current data indicates that the cosmological constant is non-zero, this situation in fact appears to be the fate of our universe in the far future, as it will be close to de Sitter. The probe particle that participates in the scattering process with the black hole needs to be some charged particle within the Standard Model (see Discussion for other possibilities); the optimal choice is the lightest charged particle.
Next, we show how this leads to a relation between the cosmological constant and parameters of the Standard Model. 

\section{Relation Between Constants}\label{Relation}
%{\em Relation Between Constants}.---

Suppose a black hole has the minimum charge $e$. We would like to be able to detect this charge from scattering the black hole off a probe. The optimal probe within the Standard Model is an electron (or positron), as it is the lightest particle of charge $e$, i.e., it has the {\em largest} charge to mass ratio of any particle or collection of particles. 
%A heavier charged particle, such as the proton, would would not be the optimal probe of the black hole's charge.
Hence the value of the electron's mass $m_e$, being the mass of the lightest charged particle, will play an important role in our results.

\begin{figure}[t]
\centering
\includegraphics[width=1\columnwidth]{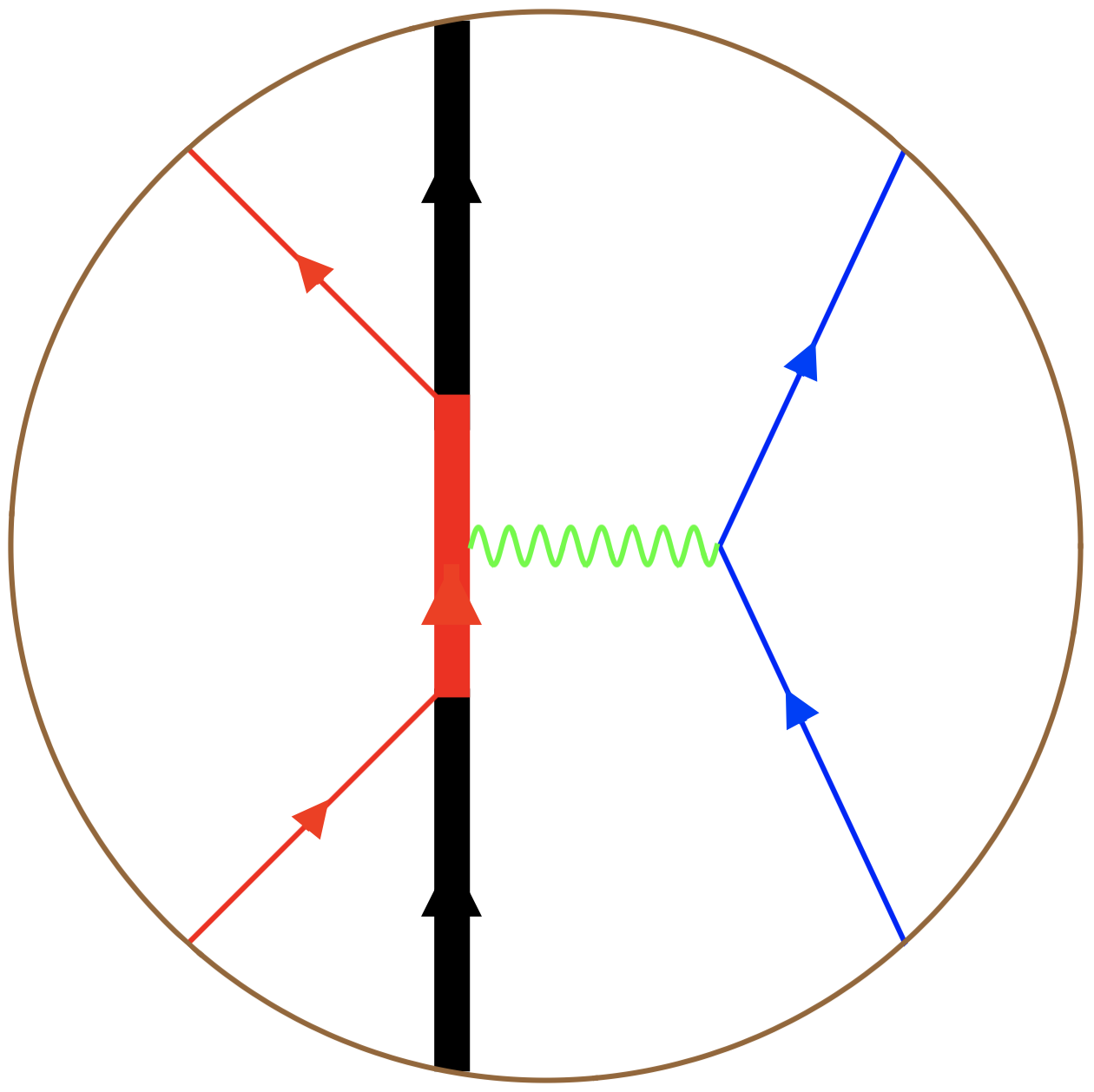}
\caption{The basic scattering process: A minimally charged black hole (thick red) is ``measured" by a probe electron (thin blue) via the electric interaction (green virtual photon exchange) if it is greater than the gravitational interaction (virtual graviton exchange). This all takes place within a de Sitter Hubble patch (brown circle). Depending on parameters, the black hole may begin neutral (thick black), absorb, then re-emit its charge (thin red) during the full process.}
\label{ScatteringDiagram} 
\end{figure}

In order for the measurement to be sharp and hence the black hole's charge is well defined in a quantum theory, we anticipate that the mutual electric potential between the black hole and the electron is comparable to, or a little larger than, their mutual gravitational potential. This can only occur for a critical black hole mass $M_*$ that obeys (units $\hbar=c=1$),
\beq
V_E={\alpha\over r}=\bb\,V_G=\bb\,{G M_* m_e\over r},
\eeq
where $\bb$ is a fudge factor, whose precise value would depend on the details of a microscopic implementation of this framework. One anticipates that $\bb$ is perhaps order 1 or a few, though $\bb=\mathcal{O}(10)$ is plausible too, as the ability to detect the black holes' charge sharply may require this. 
The scattered probing electron can be semi-relativistic, and hence its energy is on the order $m_e$.
Hence, the critical black hole mass is 
\beq
M_*={\alpha\over G \, m_e\,\bb},
\eeq
which is a mass purely specified by fundamental constants. 

Note that for sufficiently small masses, the black hole is anticipated to emit its charge via the Schwinger process \cite{Schwinger:1951nm}. At the horizon of the black hole, the electric force on an electron is $eE=\alpha/R_S^2=\alpha/(2GM_*)^2$. The standard condition for suppression of Schwinger pair production is $\pi m_e^2/(e E)\gg 1$, but here we have $\pi m_e^2/(e E) = 4\pi \alpha/b^2$, which can be smaller than 1 for plausible parameters. From a produced electron-positron pair, the oppositely charged particle can fall in and neutralize the black hole, while the other can escape. For work on neutralization rates of black holes, see Ref.~\cite{Hiscock:1990ex}. In this case, the appropriate full scattering process involves black hole charge absorption, then detection, then Schwinger emission; as depicted in Figure \ref{ScatteringDiagram}.

For detectability, we require that such a critical black hole (which may be neutral for much of the process) can survive the scattering process from the horizon and back out to the horizon -- a time of order $2\,\tH$ (where $\tH$ is the Hubble time) -- before appreciably Hawking evaporating away. 
We recall that the time for a black hole of initial mass $M$ to Hawking evaporate away completely is \cite{Hawking:1975vcx}
\beq
%\tBH={5120\pi\,G^2 M^3\over \hbar\,c^4}
\tBH(M)=5120\,\pi\,G^2 M^3.
\eeq
Also, recall that in de Sitter space, the Hubble time $\tH$ is determined by the cosmological constant $\Lambda$ (we shall use the notation that $\Lambda$ has units of energy density and the traditionally defined cosmological constant is actually $8\pi G\,\Lambda$). According to the Friedmann equation (in spatially flat slicing), we have
\beq
\tH=\Hub^{-1}=\sqrt{3\over 8\pi G\,\Lambda}.
\eeq

Altogether, we impose the inequality
\beq
\tBH(M_*)\geq 2\,\gam\,\tH,
\eeq
where $\gam$ is a fudge factor, which captures the fact that the microscopic theory may require the black hole lifetime to be somewhat larger than Hubble in order for the state of the black hole to be well defined. One may anticipate $\gam=\mathcal{O}(10)$ or so.

Solving for $\Lambda$, we obtain
\beq
\Lambda\geq\Lambda_c,
\eeq
where the critical value is found to be
\beq
%\Lambda\geq 0.03\left(\bb\over8\right)^{\!6}\left(\gam\over8\right)^{\!2} {G\,m_e^6\over\alpha^6}
\log\!\left[{\Lambda_c\over\mpl^4}\right]= -0.73+\log\!\left[G^3\,m_e^6\over\alpha^6\right]+6\log[{\bb\over10}]+2\log[{\gam\over10}],
\label{Bound}\eeq
where $\log$ means log base 10. Here we have rescaled factors $\bb$ and $\gam$ by representative factors of 10 for convenience, and we have expressed $\Lambda$ in units of the Planck mass $\mpl=1/\sqrt{G}$.

The question is whether this result is consistent with observations. 
By inserting the Standard Model measured values $G\,m_e^2\approx1.75\times10^{-45}$ and $\alpha\approx1/137$, we find the prediction
\beq
%\Lambda\geq 10^{-123}\mpl^4\left(\bb\over8\right)^{\!6}\left(\gam\over8\right)^{\!2}
\log\!\left[\Lambda_c\over\mpl^4\right]= -122.2+6\log[{\bb\over10}]+2\log[{\gam\over10}].
\eeq
Hence, if the microscopic theory gives $\bb$ and $\gam$ values $\mathcal{O}(10)$, which is plausible, and if the bound is saturated, then we obtain the prediction of an appropriately small $\Lambda$. For $\bb,\,\gam$ between 4 and 15, say, the prediction, with error bars, is $\Lambda\sim 10^{-123\pm2}\,\mpl^4$. Moreover, if we have $\bb^{3/4}\gam^{1/4}\approx8$, then we have $\Lambda\approx 10^{-123}\,\mpl^4$, which is the observed value of the density of dark energy that is driving cosmic acceleration when interpreted as a cosmological constant. The observed value is obtained from $\Lambda_{\mbox{\tiny{obs}}}=3H_0^2\,\Omega_\Lambda/(8\pi G)$ with $H_0\approx 70$\,km/s/Mpc and $\Omega_\Lambda\approx 0.7$ \cite{Planck:2018vyg}.

Saturating the above bound to give $\Lambda\approx\Lambda_c$ may be possible in a refined version of our argument. As mentioned in the Introduction, an entropic argument is a possibility: We note that if one considers the de Sitter entropy $S_{\mbox{\tiny{dS}}}=A_H/(4G)$, where $A_H=4\pi/\Hub^2$ is the horizon area \cite{Gibbons:1977mu}, then $S_{\mbox{\tiny{dS}}}=3\mpl^4/(8\Lambda)$.  Since the number of microstates is $\Omega_{\mbox{\tiny{dS}}}\sim \exp(S_{\mbox{\tiny{dS}}})$, then there is an exponential favorability of asymptotically small $\Lambda$ \cite{Hawking:1984hk}; plausibly saturating the bound.

\vspace{0.2cm}
\section{Discussion}\label{Discussion}
%{\em Discussion}.---
In this paper, we have introduced some hypothesized rules that quantum gravity may impose on charged black holes in de Sitter space.\footnote{Some other interesting works considering charged black holes in de Sitter space, includes Refs.~\cite{Montero:2019ekk,Montero:2021otb,Jiang:2005xb,Simovic:2018tdy,Mascher:2022pku}.} This led to a relation between the cosmological constant and the electron's mass and charge.
Important future work requires embedding the above principles into a microscopic framework, and checking if indeed the combination of $\bb^{3/4}\gam^{1/4}=\mathcal{O}(10)$ emerges. 

One way to falsify the proposal would be if there are dark sectors (perhaps related to dark matter) charged under some dark $U(1)_{D}$. By the same principles proposed here, one should be able to detect the black holes of this new type of discrete charge too. If the lightest charged particles $m_D$ in such a sector have masses significantly larger than the (visible) electron mass, then the strictest bound would arise from that sector. We should use Eq.~(\ref{Bound}) with $m_e\to m_D$ and $\alpha\to\alpha_D$. For sufficiently large $m_D$ (or sufficiently small $\alpha_D$) the value of $\Lambda_c$ would be larger than the observed value of dark energy and the proposal would be falsified. Such dark sectors could in principle be detected experimentally due to small mixings with the Standard Model or potentially through novel astrophysical properties if related to dark matter. 

As another possibility, if we were to observe that dark energy is relaxing over time to significantly smaller values (such as in models of quintessence \cite{Ratra:1987rm}), then again the proposal would be falsified.

\bigskip

{\bf Comment}: After this work, an article appeared \cite{Hod:2023jfo} emphasizing the role of the Schwinger effect. However, as mentioned in Sections \ref{Principles}, \ref{Relation}, this effect is accounted for here, without changing the conclusions.

\section*{Acknowledgments}
%{\em Acknowledgments}.---
We thank Miguel Montero for very helpful discussion.
We acknowledge discussion with Shahar Hod.
M.~P.~H.~ is supported in part by National Science Foundation grant PHY-2013953. 
A.~L.~ was supported in part by the Black Hole Initiative at Harvard University which is funded by grants from the John Templeton Foundation and the Gordon and Betty Moore Foundation.

%\appendix

\end{document}